\documentclass[aps,pra,preprint,superscriptaddress]{revtex4-1}
\usepackage{xcolor}
\usepackage{graphicx}
\usepackage[english]{babel}
\usepackage[utf8]{inputenc}
\usepackage{graphicx}
\usepackage{graphics}
\usepackage{braket}
\usepackage{graphicx}
\usepackage[parfill]{parskip}
\usepackage{array}
\usepackage{fancyhdr}
\usepackage{fullpage}
\usepackage{braket}
\usepackage{tabulary}
\usepackage{setspace}
\usepackage{bbold}
\usepackage{amsmath}
\usepackage{dsfont}
\DeclareMathOperator{\Tr}{Tr}
\usepackage{physics}
\RequirePackage{hyperref}
\hypersetup{
    colorlinks=true,
    linkcolor=blue,
    citecolor=blue,
    urlcolor=blue,
}

\begin{document}


\title{Using three-partite GHZ states for partial quantum error-detection in entanglement-based protocols}


\author{M. G. M. Moreno}
\email{marcosgeorge.mmf@gmail.com}
\author{Alejandro Fonseca}%
\affiliation{Departamento de Física, Universidade Federal de Pernambuco. Recife 50670-901, Pernambuco, Brazil.}
\author{Márcio M. Cunha}
\affiliation{Departamento de Matemática, Universidade Federal de Pernambuco. Recife 50670-901, Pernambuco, Brazil.}
\affiliation{Departamento de Física, Universidade Federal Rural de Pernambuco. Recife 52171-900, Pernambuco, Brazil.}


\date{\today}

\begin{abstract}
The problem of noise incidence on qubits taking part of bipartite entanglement-based protocols is addressed. It is shown that the use of a three-partite GHZ state and 
measurements instead of their EPR counterparts allows the experimenter to detect $2/3$ of the times whenever one of the qubits involved in the measurement is 
affected by bit-flip noise through the mere observation of unexpected outcomes in the teleportation and superdense coding protocols when compared to the ideal case. 
It is shown that the use of post-selection after the detection of noise leads to an enhancement in the efficiency of the protocols. The idea is extended to any protocol 
using entangled states and measurements. Furthermore it is provided a generalization in which GHZ states and measurements with an arbitrary amount of qubits are 
used instead of EPR pairs, and remarkably, it is concluded that the optimal number of qubits is only three.
\end{abstract}

\pacs{}

\maketitle
\section{Introduction}

Although it has been just over two decades since the appearance of error-correction schemes for quantum systems first introduced in a seminal paper by Shor \cite{shor1995scheme} and further extended one year later by Steane \cite{Steane96}, nowadays these protocols represent a cornerstone in quantum information science (QIS) due to the role played towards the possibility of building quantum devices large enough to be able to improve processing capacity and information storage stability when compared to classical counterparts \cite{Tehral15}. Shor's work inspired several theoretical extensions and experimental realizations, and today represents a very active area in quantum information. For a deeper exploration and recent progresses, we refer the reader to \cite{Tehral15,devitt2013quantum,Raussendor2012}.

One of the essential elements in quantum error correction codes (QEC) and hence, on the feasibility of quantum computation is \textit{entanglement} \cite{horodecki2009quantum}. Besides QEC applications, entanglement is also a key resource for a large variety of tasks in QIS \cite{vedral1998entanglement,brandao2008entanglement}, among the most known we find: Ekert's quantum key distribution \cite{ekert1991quantum}, superdense coding \cite{bennett1992communication} and the teleportation protocol \cite{bennett1993teleporting}. In fact an entire quantum computer can be conceived where entanglement provides all the basic structure \cite{raussendorf2001one}. 

Due to the importance of these protocols, several efforts have been put towards their implementation under more realistic frameworks, i.e. considering the effect of interactions with the environment. In reference \cite{taketani2012optimal} the action of noise in the teleportation protocol is contemplated and an optimal protocol is derived. A set of strategies to improve the fidelity in quantum teleportation under different kinds of noise is proposed in \cite{fortes2015fighting}. In addition, several schemes comparing multipartite channels were considered in \cite{jung2008greenberger}.

In the present work it is formulated a scheme for partial error-detection concerning protocols based on bipartite entanglement between qubits. We show that this is possible by literally replacing EPR states by tripartite Greenberger-Horne-Zeilinger (GHZ) states \cite{greenberger1989going,greenberger1990bell}, and whenever needed, instead of EPR measurements we use the GHZ basis. Our procedure, inspired by some results reported in \cite{cunha2017non}, follows the basic ideas of reference \cite{grassl1997codes}, using an ancillary system which allows for detection of the noise incidence and post-selection of the desired outputs, a process that has been considered as a viable mean of computation \cite{knill2005quantum}. On one hand our protocol is in a certain way limited, for it only permits a partial detection of one kind of noise, say bit-flip, on the other hand it is far less expensive in terms of resources: while the QEC proposed by Shor \cite{shor1995scheme} demands two extra qubits to detect bit-flip in each qubit of memory, our proposal demands only one to reveal noise on two qubits. Furthermore, our process does not demand an adjacent computation to be implemented, instead it is only necessary to adjust some steps of the existing task.

The paper is organized as follows: First we describe the approach used to model the effect of noise on the system in terms of the Kraus operators, and the domain of validity to the model. In section \ref{Noise} we compare some protocols using EPR states to the case when these are replaced by GHZ counterparts. In both cases we consider perfect realization (perfect in the sense of production of states and completion of ideal measurements) and the presence of noise. In section \ref{GP} the ideas exposed previously are extended in order to cover any protocol using pairs of entangled qubits and EPR measurements. Section \ref{NP} is devoted to show the optimality of the protocol for the case of $N=3$. In the last section the main results are discussed and some conclusions are given.

\section{Bit flip noise}
\label{Noise}

In general terms it is possible to include the noise effect on a quantum system by employing several approaches. In this paper we are not interested in the dynamics of the system in a detailed way,
thus we can use the formalism of Kraus operators \cite{Nielsen10}. This approach provides a practical way to describe several types of errors that may take place during the experimental implementation of quantum protocols. In this paper our main concern is the study of a system affected by bit-flip noise, under this, a qubit initially prepared in a state $\ket{j}$ is modified as:
\begin{equation*}
\ket{j} \rightarrow \ket{j\oplus 1}
\end{equation*}
where the symbol ``$\oplus$" indicates sum modulo $2$. 
Given a $N$-partite system, if the $k$-th qubit may with probability $p$ be affected, the Kraus operators read:
\begin{equation}
\nonumber
\hat{A}_0= \sqrt{1-p}\,\hat{\mathbb{1}}_1\otimes \cdots \otimes\hat{\mathbb{1}}_N, \hspace{0.4cm} \hat{A}_1= \sqrt{p}\,\hat{\mathbb{1}}_1\otimes \cdots \otimes\hat{\sigma}_x^{(k)}\otimes \cdots \otimes\hat{\mathbb{1}}_N.
\end{equation}
In the Kraus formalism, the quantum state evolution after the interaction with the environment may be described as a map:
\begin{equation}
\hat{\rho}\in\mathcal{B}(\mathcal{H}) \longmapsto \varepsilon(\hat{\rho})\in\mathcal{B}(\mathcal{H}),
\end{equation}
where $\mathcal{B(H)}$ is the space of the bounded operators on the Hilbert space $\mathcal{H}$. More explicitly, $\varepsilon(\hat{\rho})$ is given by:
\begin{equation}
\varepsilon (\hat{\rho})=\sum_j \hat{A}_j \hat{\rho} \hat{A}_j^{\dagger},
\end{equation}
in this way, when a single qubit described by $\hat{\rho}=\dyad{\phi}$ is affected by bit-flip noise, we have
\begin{equation}
\varepsilon (\hat{\rho})=(1-p)\dyad{\phi}+p\hat{\sigma}_x\dyad{\phi}\hat{\sigma}_x^{\dagger}.
\end{equation}
%
%
%
%
%
%
For a composed system of $N$ qubits subject to bit-flip noise acting locally in each subsystem, we have:
\begin{equation}
\label{MapN}
\varepsilon (\hat{\rho})=\sum_{j_1,\dots,j_N=0}^1 \hat{A}_{j_1}\otimes\cdots\otimes\hat{A}_{j_N} ~ \hat{\rho} ~ \hat{A}_{j_1}^{\dagger}\otimes\cdots\otimes\hat{A}_{j_N}^{\dagger}.
\end{equation}
%
%
%
%
Let us assume that every part of the system may be affected with equal probability $p$, and moreover we restrict ourselves to the weak-noise regime, i.e. the probability is low enough in order to ensure that events in which we have at most two qubits affected are very unlikely compared to those where there is only one. In this way, after some calculations, equation \ref{MapN} is reduced to:
\begin{multline}
\varepsilon(\hat{\rho})\approx(1-Np)\hat{\rho}+p\Big\{\left(\hat{\sigma}_x\otimes\mathds{\hat{1}}\otimes\cdots\otimes\mathds{\hat{1}}\right)\hat{\rho}\left(\hat{\sigma}_x\otimes\mathds{\hat{1}}\otimes\cdots\otimes\mathds{\hat{1}}\right)+\\ +\left(\mathds{\hat{1}}\otimes\hat{\sigma}_x\otimes\cdots\otimes\mathds{\hat{1}}\right)\hat{\rho}\left(\mathds{\hat{1}}\otimes\hat{\sigma}_x\otimes\cdots\otimes\mathds{\hat{1}}\right)+ ~~\cdots~~+
\\ 
+\left(\mathds{\hat{1}}\otimes\mathds{\hat{1}}\otimes\cdots\otimes\hat{\sigma}_x\right)\hat{\rho}\left(\mathds{\hat{1}}\otimes\mathds{\hat{1}}\otimes\cdots\otimes\hat{\sigma}_x\right)
\Big\}+\mathcal{O}(p^2)f(\hat{\rho}),
\end{multline}
where $f(\hat{\rho})$ represents higher order perturbations on the initially prepared state $\hat{\rho}$.

\section{Comparison between protocols}

In this section we provide a comparative overview between two very important and well known protocols in QIS: (i) quantum teleportation and (ii) superdense coding. We present the protocols in two scenarios: The first one corresponding to the traditional way, using EPR pairs weakly subject to bit-flip noise and EPR measurements. In the second scenario 
we replace all EPR states present in the system by noisy three-qubit GHZ states, and GHZ measurements.

\subsection{Quantum Teleportation}

Proposed initially by Bennet and collaborators \cite{bennett1993teleporting} and posteriorly experimentally implemented \cite{bouwmeester1997experimental}, the quantum teleportation protocol represents a very important subject because it presents how quantum mechanics can be used to develop new types of communications technologies \cite{pirandola2015advances}, and remarkably, in recent times two realizations that make the protocol feasible in the context of global communications \cite{yin2017satellite,ren2017ground} have been reported.

Let start analyzing the traditional scenario using EPR pairs, and in the following we consider the teleportation scheme using a GHZ state as the channel.

\subsubsection{Teleportation using EPR states and measurements}

Let define the EPR basis $\{\ket{\psi_{mn}}\}$, whose elements are given by:
\begin{equation}
\ket{\psi_{mn}}=\frac{1}{\sqrt{2}}\sum_{j=0}^1(-1)^{mj}\ket{j,j\oplus n},
\end{equation}
where $m,n\in\{0,1\}$. In the same way the projector of the $(m,n)$ EPR state is given by:
\begin{equation}
\hat{\Pi}_{mn}\equiv\dyad{\psi_{mn}}{\psi_{mn}}.
\end{equation}
It is worth mentioning that this set form a complete basis for the two qubit Hilbert space (also known as Bell basis) and any element $\ket{\psi_{jk}}$ may be obtained by application of Pauli matrices on the state $\ket{\psi_{00}}$:
\begin{equation}
\ket{\psi_{mn}}=\left(\hat{\sigma}_z^m\otimes\hat{\sigma}_x^n\right)\ket{\psi_{00}},
\end{equation}
where $\hat{\sigma}_{\mu}^k$ indicates $k$ times the ``$\mu$" Pauli matrix.

The goal of the teleportation protocol is to virtually send the \textit{a priori} unknown state $|\Psi\rangle=\alpha_0|0\rangle+\alpha_1|1\rangle$ from one part, let us say Alice whose qubit's states lie on the Hilbert space $\mathcal{H}_A$, to a distant part, hereafter Bob, possessing a qubit on the Hilbert space $\mathcal{H}_B$. Initially Alice and Bob share an EPR state, thus the total quantum state of the system is described by:
\begin{equation}
\nonumber
\hat{\rho}_o=\dyad{\Psi}{\Psi}_A\otimes\dyad{\psi_{00}}{\psi_{00}}_{AB},
\end{equation}
\begin{equation}
\nonumber
\hat{\rho}_o=\frac{1}{2}\sum_{jkmn=0}^1\alpha_j\alpha_k^*\dyad{jm}{kn}_A\otimes\dyad{m}{n}_B.
\end{equation}

Decomposing the Alice's part in the Bell basis, using the relation between the computational and Bell basis $\ket{mn}=\sum_k(-1)^{km}\ket{\psi_{k,m\oplus n}}/\sqrt{2}$ and after some calculations, it is possible to show that:
\begin{equation}
\hat{\rho}_o=\frac{1}{4}\sum_{mn=0}^1\hat{\Pi}_{mn}^{(A)}\otimes\Big(\hat{\sigma}_x^n\hat{\sigma}_z^m\dyad{\Psi}{\Psi}_B\hat{\sigma}_z^m\hat{\sigma}_x^n\Big)+\hat{\rho}_{null},
\end{equation}
$\hat{\rho}_{null}$ corresponds to the non-diagonal part of $\hat{\rho}_o$ (i.e. terms proportional to $\dyad{\psi_{kl}}{\psi_{mn}}$ with $k\neq m$ and $l\neq n$ on the Alice's part of the system). In forthcoming decompositions we use the same notation.

In the next step Alice performs a projective measurement in the EPR basis and according to her output, she tells Bob how to adjust his state with one out of the set of operations $\left\lbrace \mathds{1},\sigma_x,\sigma_z,\sigma_z\sigma_x\right\rbrace$. We can represent the teleportation protocol by the transformation $T:\mathcal{B(H_A)}\otimes\mathcal{B(H_B)}\mapsto\mathcal{B(H_B)}$, where $\mathcal{B(H_B)}$ is the space of the bounded operators on $\mathcal{H_B}$:
\begin{equation}
T(\hat{C})=\Tr_A\left\{\sum_{mn=0}^1\left(\hat{\Pi}_{mn}\otimes\hat{\sigma}_z^{m}\hat{\sigma}_x^{n}\right)\hat{C}\left(\hat{\Pi}_{mn}\otimes\hat{\sigma}_x^{n}\hat{\sigma}_z^{m}\right)\right\},
\end{equation}
where $\Tr_A$ represents the partial trace over Alice's part. It is straightforward to show that the application of the transformation $T$ on the initial state $\hat{\rho}_0$ gives:
\begin{equation}
T(\hat{\rho}_o)=\dyad{\Psi}{\Psi}_B,
\end{equation}
as required by the teleportation protocol.

Now we assume that before the teleportation process, a small amount of bit-flip noise affects Alice's part. 
Thus the state of the system (Alice + Bob) is modified as $\hat{\rho}_0\to\hat{\varrho}$:
\begin{equation}
\nonumber
\hat{\varrho}=(1-2p)\hat{\rho}_o+p\Big\{\left(\hat{\sigma}_x\otimes\mathds{\hat{1}}\otimes\mathds{\hat{1}}\right)\hat{\rho}_o\left(\hat{\sigma}_x\otimes\mathds{\hat{1}}\otimes\mathds{\hat{1}}\right)+\left(\mathds{\hat{1}}\otimes\hat{\sigma}_x\otimes\mathds{\hat{1}}\right)\hat{\rho}_o\left(\mathds{\hat{1}}\otimes\hat{\sigma}_x\otimes\mathds{\hat{1}}\right)\Big\}+\mathcal{O}(p^2)f(\hat{\rho}_o)
\end{equation}
where $f(\hat{\rho}_o)$ represents an operation on the initial state $\hat{\rho}_o$ which is not interesting under the scope of the present work. By noticing that $\hat{\sigma}_x^{k}\otimes\hat{\mathds{1}}\ket{\psi_{mn}}=(-1)^{km}\ket{\psi_{m,n\oplus k}}$ and $\hat{\mathds{1}}\otimes\hat{\sigma}_x^{k}\ket{\psi_{mn}}=\ket{\psi_{m,n\oplus k}}$, and after some calculations the density operator reduces to:  
\begin{equation}
\hat{\varrho}=(1-2p)\hat{\rho}_o + \frac{p}{2}\sum_{mn=0}^1\hat{\Pi}_{m,n\oplus 1}^{(A)}\otimes\Big(\hat{\sigma}_x^n\hat{\sigma}_z^m\dyad{\Psi}{\Psi}_B\hat{\sigma}_z^m\hat{\sigma}_x^n\Big)+\hat{\varrho}_{null}.
\end{equation}
%
%
%

Note above that the effect of the noise in the final state is to produce an extra term where the adjustment to be performed according to the output of the measurement is not the correct one. Proceeding with the protocol will have:
\begin{equation}
\label{telEPRfinal}
T(\hat{\varrho})=(1-2p)\dyad{\Psi}{\Psi}+2p(\hat{\sigma}_x\dyad{\Psi}{\Psi}\hat{\sigma}_x).
\end{equation}
Hence only with probability $(1-2p)$ Bob's part holds in the desired state. 

Now we turn our attention to the teleportation protocol but instead of EPR states, in this case we use three partite GHZ states and measurements.

\subsubsection{Teleportation using GHZ states and measurements}

First let us define the three partite GHZ basis elements and related projectors:
\begin{equation}
\ket{\phi_{kmn}}=\frac{1}{\sqrt{2}}\sum_{j=0}^1(-1)^{kj}\ket{j,j\oplus m,j\oplus n},
\end{equation}
and
\begin{equation}
\hat{\Pi}_{kmn}'\equiv\dyad{\phi_{kmn}}{\phi_{kmn}},
\end{equation}
where $k,m,n\in\{0,1\}$. Analogously to the EPR basis, any element $\ket{\phi_{kmn}}$ may be obtained by application of Pauli matrices on the state $\ket{\phi_{000}}=\left(\ket{000}+\ket{111}\right)/\sqrt{2}$:
\begin{equation}
\ket{\phi_{kmn}}=\left(\hat{\sigma}_z^k\otimes\hat{\sigma}_x^m\otimes\hat{\sigma}_x^n\right)\ket{\phi_{000}}.
\end{equation}
%
%
%

Now Alice and Bob share three entangled qubits prepared in a GHZ state $\ket{\phi_{000}}$, two qubits are held by Alice and the other one by Bob. Alice's objective is again to send an unknown qubit state $\ket{\Psi}$ to Bob. The initial state of the four qubits reads:
\begin{equation}
\nonumber
\hat{\rho}_o'=\dyad{\Psi}{\Psi}_A\otimes\dyad{\phi_{000}}{\phi_{000}}_{AAB},
\end{equation}
\begin{equation}
\label{telrhoGHZ}
\hat{\rho}_o'=\frac{1}{2}\sum_{jkmn=0}^1\alpha_j\alpha_k^*\dyad{jmm}{knn}_A\otimes\dyad{m}{n}_B.
\end{equation}

We can use the relation $\ket{kmn}=\sum_j(-1)^{jk}\ket{\phi_{j,k\oplus m,k\oplus n}}/\sqrt{2}$ to rewrite equation \ref{telrhoGHZ} in the following way:
\begin{equation}
\hat{\rho}_o'=\frac{1}{4}\sum_{mn=0}^1\hat{\Pi}_{mnn}'^{(A)}\otimes\Big(\hat{\sigma}_x^n\hat{\sigma}_z^m\dyad{\Psi}{\Psi}_B\hat{\sigma}_z^m\hat{\sigma}_x^n\Big)+\hat{\rho}_{null}'.
\end{equation}
%
From the projector on Alice's part we can see that instead of eight, there are only four possible outputs for a measurement in the GHZ basis, and the desired operation to accomplish the teleportation process can be described by:
\begin{equation}
T'(\hat{C})=\Tr_A\left\{\sum_{mn=0}^1\left(\hat{\Pi}_{mnn}'\otimes\hat{\sigma}_z^{m}\hat{\sigma}_x^{n}\right)\hat{C}\left(\hat{\Pi}_{mnn}'\otimes\hat{\sigma}_x^{n}\hat{\sigma}_z^{m}\right)\right\},
\end{equation}
%
%
It is straightforward to show that $T'(\hat{\rho}_o')=\dyad{\Psi}{\Psi}_B$, as expected.

As in the previous case, before the teleportation protocol takes process, we consider bit-flip noise acting on all Alice's qubits under a weak regime. Thus we have $\hat{\rho}_o'\to\hat{\varrho}'$:
\begin{multline}
\nonumber
\hat{\varrho}'=(1-3p)\hat{\rho}_o'+p\Big\{\left(\hat{\sigma}_x\otimes\mathds{\hat{1}}\otimes\mathds{\hat{1}}\right)\hat{\rho}_o'\left(\hat{\sigma}_x\otimes\mathds{\hat{1}}\otimes\mathds{\hat{1}}\right)+\left(\mathds{\hat{1}}\otimes\hat{\sigma}_x\otimes\mathds{\hat{1}}\right)\hat{\rho}_o'\left(\mathds{\hat{1}}\otimes\hat{\sigma}_x\otimes\mathds{\hat{1}}\right)+
\\ 
+\left(\mathds{\hat{1}}\otimes\mathds{\hat{1}}\otimes\hat{\sigma}_x\right)\hat{\rho}_o'\left(\mathds{\hat{1}}\otimes\mathds{\hat{1}}\otimes\hat{\sigma}_x\right)
\Big\}+\mathcal{O}(p^2)f'(\hat{\rho}_o').
\end{multline}
In this case we take into account the following relations: $\hat{\sigma}_x^{j}\otimes\hat{\mathds{1}}\otimes\hat{\mathds{1}}\ket{\phi_{k,m,n}}=(-1)^{jk}\ket{\phi_{k,m\oplus j,n\oplus j}}$, $\hat{\mathds{1}}\otimes\hat{\sigma}_x^{j}\otimes\hat{\mathds{1}}\ket{\phi_{k,m,n}}=\ket{\phi_{k,m\oplus j,n}}$ and $\hat{\mathds{1}}\otimes\hat{\mathds{1}}\otimes\hat{\sigma}_x^{j}\ket{\phi_{k,m,n}}=\ket{\phi_{k,m,n\oplus j}}$.
Substituting, we have:
%
%
\begin{equation}
\nonumber
\hat{\varrho}'=(1-3p)\hat{\rho}_o'+\frac{p}{4}\sum_{mn=0}^1\left(\hat{\Pi}'^{(A)}_{m,n\oplus 1,n\oplus 1}+\hat{\Pi}'^{(A)}_{m,n\oplus 1,n}+\hat{\Pi}'^{(A)}_{m,n,n\oplus 1}\right)\otimes\Big(\hat{\sigma}_x^n\hat{\sigma}_z^m\dyad{\Psi}{\Psi}_B\hat{\sigma}_z^m\hat{\sigma}_x^n\Big)+\hat{\varrho}_{null}'.
\end{equation}
%
%
%
Here we can find a new element: while in the ideal case we would only get four outputs, due to the incidence of noise we have additionally four measurement results, say: $\left\lbrace\ket{\phi_{001}},\ket{\phi_{010}},\ket{\phi_{101}},\ket{\phi_{110}}\right\rbrace$ and even though we cannot proceed as in the traditional error-syndrome, because here we can only determine the cases in which noise has affected one of the GHZ qubits in Alice, but we are not capable of identify on which specifically. However we can take advantage of this partial knowledge to perform a post-selection during the protocol: discarding Bob's qubits whenever ``undesired" outputs are obtained.


Following this process and performing the proper renormalization, the final state reads:
\begin{equation}
\hat{\varrho}_F=\frac{1-3p}{1-2p}|\Psi\rangle\langle\Psi|+\frac{p}{1-2p}\hat{\sigma}_x\dyad{\Psi}{\Psi}\hat{\sigma}_x,
\end{equation}
which is always better than the result obtained in the protocol using EPR states, (eq. \ref{telEPRfinal}) under a weak-noise regime.

\subsection{Superdense Coding}
Here we analyze the protocol of superdense coding, theoretically proposed in \cite{bennett1992communication} and experimentally implemented in \cite{mattle1996dense}, whose importance lies on its capability to reach a compression factor of $2$ per message. The scenario consists on two parts: Alice (the sender), sharing an EPR state $\ket{\psi_{00}}$ with Bob (the receiver).

Through the application of the operations $\hat{\sigma}_x^{a_2}\hat{\sigma}_z^{a_1}$, Alice encodes a two bit message $(a_1,a_2)$ on her qubit which is afterwards physically sent to Bob. The receiver (now holding both qubits) performs a measurement in the EPR basis and recovers the original message according to his output: $\ket{\psi_{a_1,a_2}}$.

If nevertheless before Bob's measurement the system is affected by bit-flip noise, then under the regime of weak noise, the resulting state is given by:
\begin{equation}
\hat{\varrho}=(1-2p)\hat{\Pi}_{a_1,a_2}+2p\hat{\Pi}_{a_1,a_2\oplus1},
\end{equation}
thus only a fraction of $1-2p$ outputs will lead to a proper interpretation of the original message. 

It is possible to think on a different procedure to carry out this protocol.
Assume that Alice and Bob share a GHZ state $\ket{\phi_{000}}_{ABB}$ instead of an EPR state. 
Alice encodes her message in the same way as before and sends her qubit to Bob who performs a measurement in the GHZ basis. He recovers Alice's string by associating the output $\ket{\phi_{k,m,n}}$ with the string $(k,m)$. Notice that, so far, the outputs with $m\neq n$ are not possible, however when the action of noise on the system is considered (before Bob's measurement), the system is described by:
\begin{equation}
\hat{\varrho}'=(1-3p)\hat{\Pi}_{a_1,a_2,a_2}'+p\left(\hat{\Pi}_{a_1,a_2\oplus 1,a_2\oplus 1}'+\hat{\Pi}_{a_1,a_2\oplus 1,a_2}'+\hat{\Pi}_{a_1,a_2,a_2\oplus 1}'\right).
\end{equation}
In this case, after performing a measurement in the GHZ basis, Bob will receive a fraction of $1-3p$ of correct signal. However he is capable of detecting that a flip has happened by observing the detection of ``disallowed outputs" ($\hat{\Pi}_{k,m,n}'$ for $m\neq n$). She can thus discard those results and increase to a rate of $(1-3p)/(1-2p)$ accurate strings, which is always better than $1-2p$ in the weak noise regime.

\section{General protocol}
\label{GP}
\subsection{Any protocol involving EPR pairs may be performed using GHZ states}
In this section our goal is to demonstrate that any protocol involving an EPR pair and measurements can be performed using a GHZ state and measurements. With this result we show that, by using GHZ configurations it is possible to detect the presence of bit-flip (or phase-flip) noise and reach a higher precision by post-selection.

Assume we are given a task to be performed using an EPR pair ($\ket{\psi_{mn}}$) and an arbitrary subsystem described by the density operator $\hat{\rho}_S\in\mathcal{B}(\mathcal{H}_S)$. The initial state of the system as a whole reads:
\begin{equation}
\hat{\rho}_o=\hat{\rho}_S\otimes\hat{\Pi}_{mn},
\end{equation}
for $\hat{\Pi}_{mn}\in\mathcal{B}(\mathcal{H}_1\otimes\mathcal{H}_2)$, with $\mathcal{H}_j$ the Hilbert space associated to the qubits of the EPR state.

In order to perform the task in question, eventually some operation $T$ will be performed on $\hat{\rho}_o$ transforming into a new state $\hat{\rho}_f$. Hence at some point we must have:
\begin{equation}
T(\hat{\rho}_o)=\hat{\rho}_f.
\end{equation}
In general such a transformation may be written as:
\begin{equation}
\label{T}
T(\hat{\sigma})=\Tr_{\mathcal{H}_Q}\left(\sum_{k=1}^{m'} \hat{V}_k \hat{\sigma} \hat{V}_k^{\dagger} \right)\Gamma_{m'}^{-1},
\end{equation}
where $\Gamma_{m'}=\Tr\left\{\sum_{k=1}^{m'} \hat{V}_k\hat{\sigma} \hat{V}_k^{\dagger}\right\}$ and $\sum_k^{m} \hat{V}_k^{\dagger}\hat{V}_k=\hat{\mathds{1}}$, for a given $m'\leq m$,
under an arbitrary Hilbert space $\mathcal{H}_Q$ satisfying: $\mathcal{H}_Q\subset\mathcal{H}_S\otimes\mathcal{H}_1\otimes\mathcal{H}_2$.

We can always add an extra qubit $\hat{\rho}_{anc}=\dyad{0}{0}\in\mathcal{B}(\mathcal{H}_{anc})$, sometimes called \textit{ancila} and define the extended state $\hat{\rho}'_o$ as follows:
\begin{equation}
\hat{\rho}'_o= (\hat{\mathds{1}}_S\otimes\hat{\mathds{1}}_1\otimes \hat{C}_{not})\hat{\rho}_o\otimes \hat{\rho}_{anc}(\hat{\mathds{1}}_S\otimes\hat{\mathds{1}}_1\otimes \hat{C}_{not}^{\dagger})=\hat{\rho}_S\otimes\hat{\Pi}_{mnn}',
\end{equation}
where $\hat{C}_{not}$ represents the CNOT operation acting on $\mathcal{H}_{2}\otimes\mathcal{H}_{anc}$.

The transformation defined previously (eq. \ref{T}) may be generalized to the extended Hilbert space by introducing the operators $\hat{V}'_k \in\mathcal{B}(\mathcal{H}_S\otimes\mathcal{H}_0\otimes\mathcal{H}_1\otimes\mathcal{H}_{anc})$ in the following way:
\begin{equation}
\hat{V}_k'=(\hat{V}_k\otimes\hat{\mathds{1}}_{anc})(\hat{\mathds{1}}_S\otimes\hat{\mathds{1}}_1\otimes \hat{C}_{not}^{\dagger}),
\end{equation}
likewise, the transformation $T'$ holds:
\begin{equation}
T'(\hat{\sigma})=\Tr_{\mathcal{H}_{Q'}}\left(\sum_{k=1}^{m'} \hat{V}_k'\hat{\sigma}\hat{V}_k'^{\dagger}\right),
\end{equation}
which implies that
\begin{equation}
T(\hat{\rho}_o) = T'(\hat{\rho}_o')=\hat{\rho}_f.
\end{equation}
In conclusion, any task making use of EPR pairs may be performed using the same number of GHZ states.

\subsection{Noise detection}

Now we focus on tasks using EPR states that involve detecting some elements of an EPR basis $\{\ket{\psi_{mn}}\}$ i.e. $T$ is of the form:
\begin{equation}
T(\hat{\rho})=\Tr_{\mathcal{H}_Q}\left\{\sum_{mn}\left(\hat{U}_{mn}\otimes\hat{\Pi}_{mn}\right)\hat{\rho}\left(\hat{U}_{mn}^{\dagger}\otimes\hat{\Pi}_{mn}\right)\right\}
\end{equation}
where $\hat{U}_{mn}$ are arbitrary unitary matrices. It is important to remark that we assume the measurement acting on at least one of the qubits in the EPR state.

Consider the incidence of bit-flip noise on the qubits involved in the measurement. Under the weak noise regime we have:
\begin{multline}
T(\hat{\varrho}_o)=(1-2p)\Tr_{\mathcal{H}_Q}\left\{\sum_{mn=0}^1\left(\hat{U}_{mn}\otimes\hat{\Pi}_{mn}\right)\hat{\rho}_o \left(\hat{U}_{mn}^{\dagger}\otimes\hat{\Pi}_{mn}\right)\right\}+\\ +2p\Tr_{\mathcal{H}_Q}\left\{\sum_{mn=0}^1\left(\hat{U}_{mn}\otimes\hat{\Pi}_{m,n\oplus 1}\right)\hat{\rho}_o\left(\hat{U}_{mn}^{\dagger}\otimes\hat{\Pi}_{m,n\oplus 1}\right)\right\},
\end{multline}
\begin{equation}
T(\hat{\varrho}_o)=(1-2p)\hat{\rho}_f+2p\Tr_{\mathcal{H}_Q}\left\{\sum_{mn=0}^1\left(\hat{U}_{mn}\otimes\hat{\Pi}_{m,n\oplus 1}\right)\hat{\rho}_o\left(\hat{U}_{mn}^{\dagger}\otimes\hat{\Pi}_{m,n\oplus 1}\right)\right\}.
\end{equation}
As we can see, the effect of bit-flip noise on the qubits involved in the measurement is to mix the labels, and in this way decrease the precision of the operation.


Following the results of the last section we are able to modify the transformation above in order to replace EPR states by GHZ states. Thus it is straightforward to show that the corresponding operation $T'$ is:
\begin{equation}
\label{T'}
T'(\hat{C})=\Tr_{\mathcal{H}_{Q'}}\left\{\sum_{mn=0}^1\left(\hat{U}_{mn}\otimes\hat{\Pi}_{mnn}'\right)\hat{C}\left(\hat{U}_{mn}^{\dagger}\otimes\hat{\Pi}_{mnn}'\right)\right\}.
\end{equation}
On the other hand, if we look to the GHZ version of the protocol, we have:
\begin{multline}
T'(\hat{\varrho}_o')=(1-3p)\Tr_{\mathcal{H}_{Q'}}\left\{\sum_{mn=0}^1\left(\hat{U}_{mn}\otimes\hat{\Pi}_{mnn}'\right)\hat{\rho}_o'\left(\hat{U}_{mn}^{\dagger}\otimes\hat{\Pi}_{mnn}'\right)\right\}+\\
+p\Tr_{\mathcal{H}_{Q'}}\left\{\sum_{mn=0}^1\left(\hat{U}_{mn}\otimes\hat{\Pi}_{m,n\oplus1,n\oplus1}'\right)\hat{\rho}_o'\left(\hat{U}_{mn}^{\dagger}\otimes\hat{\Pi}_{m,n\oplus1,n\oplus1}'\right)\right\}+\\
+p\Tr_{\mathcal{H}_{Q'}}\left\{\sum_{mn=0}^1\left(\hat{U}_{mn}\otimes\hat{\Pi}_{m,n\oplus1,n}'\right)\hat{\rho}_o'\left(\hat{U}_{mn}^{\dagger}\otimes\hat{\Pi}_{m,n\oplus1,n}'\right)\right\}+\\
+p\Tr_{\mathcal{H}_{Q'}}\left\{\sum_{mn=0}^1\left(\hat{U}_{mn}\otimes\hat{\Pi}_{m,n,n\oplus1}'\right)\hat{\rho}_o'\left(\hat{U}_{mn}^{\dagger}\otimes\hat{\Pi}_{m,n,n\oplus1}'\right)\right\}.
\end{multline}

It is clear by the above equation that the effect of noise in this scenario is more rich. We observe as in the EPR protocol a mixing of the detections labels, however it happens in a smaller proportion, and to compensate that effect, new outcomes become possible. As those did not suppose to appear, we can always perform a post-selection by eliminating those outcomes. This can be done in the generic operation $\tilde{T}'$:
\begin{multline}
\tilde{T}'(\hat{\varrho}_o')=\Gamma\cdot\Bigg\{(1-3p)\Tr_{\mathcal{H}_{Q'}}\left[\sum_{mn=0}^1\left(\hat{U}_{mn}\otimes\hat{\Pi}_{mnn}'\right)\hat{\rho}_o'\left(\hat{U}_{mn}^{\dagger}\otimes\hat{\Pi}_{mnn}'\right)\right]+\\
+p\Tr_{\mathcal{H}_{Q'}}\left[\sum_{mn=0}^1\left(\hat{U}_{mn}\otimes\hat{\Pi}_{m,n\oplus1,n\oplus1}'\right)\hat{\rho}_o'\left(\hat{U}_{mn}^{\dagger}\otimes\hat{\Pi}_{m,n\oplus1,n\oplus1}'\right)\right]\Bigg\},
\end{multline}
where $\Gamma$ is a normalization factor, depends on the unexpected new outputs and is equal to $1-2p$.
%
%

\section{General protocol using $N$-partite GHZ states}
\label{NP}

A natural point remaining to be explored in this work is whether the improvement in the protocols observed here is only a manifestation of a non-ideal version of some error-correction scheme, whose imperfection might arise from the low size of the employed ancilla (0.5 qubits of the ancilla per qubit in the original protocol). This can be verified by generalizing this protocol to the case where more than one qubit is attached to the original system, i.e., replacing EPR pairs and measurements by $N$-partite GHZ states and measurements, with $N$ arbitrary. Such a generalization is straightforward, first define the $N$-partite GHZ state as follows:
\begin{equation}
\ket{\phi_{\vec{\mu}}}=\frac{1}{\sqrt{2}}\sum_{j=0}^1(-1)^{j\mu_0}\ket{j,j\oplus\mu_1,\dots,j\oplus\mu_{N-1}},
\end{equation}
where, $\vec{\mu}=\left(\mu_0,\dots,\mu_{N-1}\right)$, with $\mu_j\in\{0,1\}$. In analogy to the previous cases, this state may also be obtained from local application of Pauli operators on the state $\ket{\phi_{0,\dots,0}}$, as:
\begin{equation}
\ket{\phi_{\vec{\mu}}}=\left(\hat{\sigma}_z^{\mu_0}\otimes\hat{\sigma}_x^{\mu_1}\otimes\cdots\otimes\hat{\sigma}_x^{\mu_{N-1}}\right)\ket{\phi_{0,\dots,0}}.
\end{equation}
In the same way, the projector reads:
\begin{equation}
\hat{\Pi}_{\vec{\mu}}''=\dyad{\phi_{\vec{\mu}}}.
\end{equation}
Thus, the operation defined in equation \ref{T'} in this case holds:
\begin{equation}
T''(\hat{C})=\Tr_{\mathcal{H}_{Q''}}\left\{\sum_{mn=0}^1\left(\hat{U}_{mn}\otimes\hat{\Pi}_{m,n,n,\dots,n}''\right)\hat{C}\left(\hat{U}_{mn}^{\dagger}\otimes\hat{\Pi}_{m,n,n,\dots,n}''\right)\right\}.
\end{equation}
In the presence of bit-flip noise, under the weak-noise regime, we have:
\begin{multline}
\tilde{T}''(\hat{\varrho}_o'')=\Gamma\cdot\Bigg\{(1-Np)\Tr_{\mathcal{H}_{Q''}}\left[\sum_{mn=0}^1\left(\hat{U}_{mn}\otimes\hat{\Pi}_{m,n,\dots,n}''\right)\hat{\rho}_o'\left(\hat{U}_{mn}^{\dagger}\otimes\hat{\Pi}_{m,n,\dots,n}''\right)\right]+\\
+p\Tr_{\mathcal{H}_{Q''}}\left[\sum_{mn=0}^1\left(\hat{U}_{mn}\otimes\hat{\Pi}_{m,n\oplus1,n\oplus1,\dots,n\oplus1}''\right)\hat{\rho}_o''\left(\hat{U}_{mn}^{\dagger}\otimes\hat{\Pi}_{m,n\oplus1,n\oplus1,\dots,n\oplus1}''\right)\right]+\\
+p\Tr_{\mathcal{H}_{Q''}}\left[\sum_{mn=0}^1\left(\hat{U}_{mn}\otimes\hat{\Pi}_{m,n\oplus1,n,\dots,n}''\right)\hat{\rho}_o''\left(\hat{U}_{mn}^{\dagger}\otimes\hat{\Pi}_{m,n\oplus1,n,\dots,n}''\right)\right]+\\
+~~~\cdots~~~+\\
+p\Tr_{\mathcal{H}_{Q''}}\left[\sum_{mn=0}^1\left(\hat{U}_{mn}\otimes\hat{\Pi}_{m,n,n,\dots,n\oplus1}''\right)\hat{\rho}_o''\left(\hat{U}_{mn}^{\dagger}\otimes\hat{\Pi}_{m,n,n,\dots,n\oplus1}''\right)\right]\Bigg\}.
\end{multline}
Therefore, after post-selection we get an efficiency of $(1-Np)/\left[1-(N-1)p\right]$. Now it is natural to ask whether increasing the number of parties above $N=3$ enhances the performance of the protocol, i.e. for what values of $N^*$ and arbitrary $p$, the following condition holds:
\begin{equation}
\nonumber
\frac{1-N^*p}{1-(N^*-1)p} >\frac{1-3p}{1-2p}, 
\end{equation}
%
however this is only possible for $N^*<3$. In this way we have shown that the best strategy is to employ three-partite GHZ states and measurements.

\section{Discussion and Conclusion}

We presented a protocol for error detection in some entanglement-based tasks, in which the key ingredient is the replacement of EPR pairs by GHZ states. First, we showed its performance in two well known tasks: teleportation and superdense coding. Then we considered a general task under the effect of weak bit-flip noise, and we have demonstrated that it is alway possible to increase its efficiency by using our protocol. Additionally, this process is less expensive than many other protocols, as it only demands one ancillary qubit per pair of qubits in the system.
Nonetheless it is important to remark that such an enhancement is not for free, in fact, measurements in the EPR basis must be replaced by measurements in the GHZ basis. Moreover, in the form we have presented, the protocol works only under bit-flip noise, and although it is possible to conceive an adjustement in order to work with phase-flip noise through a simple local change of basis, it will never work with both kinds of noise. Furthermore we have proven that our proposal is not equivalent to any QEC by showing that it cannot be improved by increasing the size of the ancilla.

It is a curious fact that the replacement of an EPR pair by a GHZ state as a resource leads to an increase in the precision of some task that may be performed between two remote parts, for instance the protocol of teleportation. That because if we consider the bipartition involved in this protocol, the three-partite GHZ and the EPR state have the same entropy of entanglement \cite{horodecki2009quantum} and thus they should represent the same resource with the same potential (which should be the maximum possible) for such a task.

An interesting question is whether a similar process can be performed for tasks demanding a higher number of entangled parts using another relevant classes of quantum states, such as cluster states, for instance.


%
\acknowledgements
M\'arcio M. Cunha is supported  by FACEPE-FULBRIGHT BCT 0060-1.05/18 grant. Financial support from Conselho Nacional de Desenvolvimento Cient\'{\i}fico e Tecnol\'ogico (CNPq) through its program INCT-IQ, Coordena\c{c}\~ao de Aperfei\c{c}oamento de Pessoal de N\'{\i}vel Superior (CAPES), and Funda\c{c}\~ao de Amparo \`a Ci\^encia e Tecnologia do Estado de Pernambuco (FACEPE) is acknowledged.
%
\bibliography{Bibliography}

\end{document}